\documentclass[12pt]{article}
\usepackage{amsmath,amsfonts}
\advance \textheight by \topskip
\makeatletter
\@addtoreset{equation}{section}
 
\makeatother

\def\NN{\mathbb N}

\def\CC{\mathbb C}
\def\RR{\mathbb R}
\newcommand{\beq}{\begin{equation}}
\newcommand{\bee}{\end{equation}}
\newcommand{\beqa}{\begin{eqnarray}}
\newcommand{\beea}{\end{eqnarray}}

\def\>{\rangle}
\def\<{\langle}
\def\I{\hbox{1\hskip -3.2pt I}}
\def\ds{\displaystyle}

\begin{document}

\title{
\begin{flushright}
{\small USACH-FM-01-02}\\[-0.4cm]
{\small PM-01-07}\\[1cm]
\end{flushright}
{\bf Fractional helicity, Lorentz symmetry
breaking, compactification and anyons }}

\author{
{\sf S. M. Klishevich}\thanks{e-mail:
sklishev@lauca.usach.cl}$\,\,$${}^{a,b},$
{\sf M. S. Plyushchay}\thanks{e-mail:
mplyushc@lauca.usach.cl}$\,\,$${}^{a,b}$
 and
{\sf M.~Rausch de Traubenberg}\thanks{e-mail:
rausch@lpm.univ-montp2.fr}$\,\,$$^{c,d}$\\
\\
{\small ${}^{a}${\it Departamento de F\'{\i}sica,
Universidad de Santiago de Chile,
Casilla 307, Santiago 2, Chile}}\\
{\small ${}^{b}${\it Institute for High Energy Physics,
Protvino,
Russia}}\\
{\small ${}^{c}${\it Laboratoire de Physique
Math\'ematique,
Universit\'e Montpellier II,}}\\
{\small {\it Place E. Bataillon, 34095 Montpellier,
France}}\\
{\small ${}^{d}${\it
Laboratoire de Physique Th\'eorique, 3 rue de
l'Universit\'e, 67084 Strasbourg, France}}}
\date{}
\maketitle
\vskip-1.5cm

\begin{abstract}
We construct the covariant, spinor sets of relativistic wave
equations for a massless field on the basis of the two
copies of the R-deformed Heisenberg algebra. For the
finite-dimensional representations of the algebra they give
a universal description of the states with integer and
half-integer helicity. The infinite-dimensional
representations correspond formally to the massless states
with fractional (real) helicity. The solutions of the latter
type, however, break down the (3+1)$D$ Poincar\'e invariance
to the (2+1)$D$ Poincar\'e invariance, and via a
compactification on a circle a consistent theory for massive
anyons in $D$=2+1 is produced. A general
analysis of the ``helicity equation'' shows that the
(3+1)$D$ Poincar\'e group has no massless irreducible
representations with the trivial non-compact part of the
little group constructed on the basis of the
infinite-dimensional representations of $sl(2,\CC)$. This
result is in contrast with the massive case where integer
and half-integer spin states can be described on the basis
of such representations, and means, in particular, that the
(3+1)$D$ Dirac positive energy covariant equations have no
massless limit.
\end{abstract}

\newpage
\section{Introduction}
It is generally accepted that the $(2+1)$-dimensional
space-time reveals specific characteristics
which are no more valid in higher dimensions.
For instance, only in (2+1)$D$ the states
of arbitrary spin $\lambda\in \RR$ and of corresponding
anyonic statistics \cite{anyon,anyon1} different from the
bosonic and fermionic ones do exist\footnote{Strictly
speaking, other generalizations of statistics called
parafermions and parabosons exist in any space-time
dimension \cite{para,para1,para2,para3,ok}, but
via the so-called Green anzatz they can be represented in
terms of ordinary bosons and fermions.}.
There are several ways to introduce anyons in
(2+1)$D$ including the Chern-Simons
\cite{cs,cs1,cs2} and the group-theoretical
\cite{an0,an0b,group,group1,group2}
approaches.
In the latter, noticing that the $SO(2,1)$ group
is infinitely connected, the
massive anyons of spin $\lambda\in\RR$
are realized on the infinite-dimensional
half-bounded representations of $\overline{SL(2,\RR)}$
\cite{lang}, the universal covering group of $SO(2,1)$.
Within that approach, in particular, the anyons
can be described by the covariant vector \cite{vect}
and spinor \cite{spin,spin1} sets of linear
differential equations for infinite-component fields.

The little group for the (2+1)$D$ massive anyon is $SO(2)$
and it carries the same number of physical degrees of
freedom as a (2+1)$D$ massive scalar field \cite{corply}.
This specific feature is also valid for the massless states
in (3+1)$D$. Indeed, in (3+1)$D$ the little group in the
massless case is $E(2)$, the group of rotations and
translations in the $2D$ Euclidean space. Representing its
non-compact part by zero, we reduce $E(2)$ to $SO(2)$. With
this observation at hands, it was shown that some (2+1)$D$
models for anyons can be obtained from the (3+1)$D$ models
of massless particles via the appropriate formal reduction
\cite{mp}. Further, since, unlike the massive $(3+1)D$ case,
the algebra of the little group gives no quantization
restrictions, it seems that there is no strong reason for
excluding the possibility to consider massless states with
fractional (arbitrary real) helicity \cite{ryder}. Following
this line of reasoning, some time ago it was claimed that,
in fact, a massless analogue of the original Dirac spinor
set of equations describing a massive spin-$0$ field in
(3+1)$D$ \cite{dirac,dirac1} gives rise to the (3+1)$D$
massless states of helicity $\pm 1/4$ called ``quartions''
\cite{vs,vs1}. Besides, it was noted that the non-covariant
formal quantization of the massless superparticle preserving
its classical $P$-invariance should result in the
supermultiplet with helicity structure ($-1/4,+1/4$)
\cite{super}. On the other hand, there exist the topological
arguments (see, e.g., \cite{y,wein}) related to the
two-connectedness of $SO(3,1)$ which restrict the helicity
of massless representations to integer and half-integer
values. But then, {\em appealing to the well-known relations
between $D$-dimensional massive and ($D$+1)-dimensional
massless representations, one can ask what corresponds to
the (2+1)$D$ massive representations with fractional spin.
Formally, it could be the (3+1)$D$ massless irreducible
representations with fractional helicity constructed on the
basis of infinite-dimensional representations of
$sl(2,\CC)$. But since, due to the topological reasons, they
cannot exist, what kind of defects have to appear in such a
theory and how the (2+1)$D$ massive anyons emerge
from the corresponding (3+1)$D$ massless theory?}

In this paper we address in detail the problem of the
description of fractional helicity massless fields in
(3+1)$D$ on the basis of the infinite-dimensional
representations of the $sl(2,\CC)$ Lie algebra realized in
terms of the two copies of the $R$-deformed Heisenberg
algebra (RDHA) \cite{ok,rdha,rdha0}.
The RDHA was first introduced by Yang \cite{yang} in the
context of Wigner generalized quantization schemes
\cite{wigner} underlying the concept of parafields and
parastatistics \cite{ok} (in this context, see also Ref.
\cite{ohnuki}). This algebra and its
generalizations is nowadays exploited extensively in the
mathematical and physical literature in different aspects
(see Refs. \cite{dunkl,spin,spin1,susy-pb,vinet}).
It should be emphasized that the infinite-dimensional
representations of $sl(2,\CC)$ we consider have nothing to
do with those representations corresponding to the little
group $E(2)$ with a non-trivial non-compact part, to which
are usually referred as to representations with
``continuous'' \cite{bw} or ``infinite'' \cite{y} spin.

We observe here that the corresponding irreducible
infinite-dimensional representations of $sl(2,\CC)$ cannot
be ``exponentiated'' to representations of the $SL(2,\CC)$
Lie group in the massless case. In other words, the
fractional helicity representation of the little group
$SO(2)$ cannot be promoted to a representation of the (3+1)$
D$ Lorentz group being a subgroup of the corresponding
massless representations of the (3+1)-dimensional Poincar\'e
group. This is reflected in breaking of the Lorentz
invariance at the level of solutions of the {\it covariant}
spinor set of equations for fractional helicity massless
fields. The symmetry breaking corresponds to the violation
of the invariance with respect to the rotations in two
directions and to the boosts in one direction. Consequently,
the Lorentz group $SL(2,\CC)$ is broken down to
$\overline{SL(2,\RR)}$, and via a compactification and
subsequent dimensional reduction from (3+1)$D$ to (2+1)$D$ a
consistent theory for massive relativistic anyons in $D=2+1$
is produced. At the same time, we show that the (3+1)$D$
Poincar\'e group has no massless irreducible representations
characterized by the trivial non-compact part of the little
group and which would be constructed on the basis of the
infinite-dimensional representations of
$sl(2,\CC)$. This results in the same restriction for
helicity but not of a topological origin.

The paper is organized as follows. In Section \ref{rdha}
we realize representation of $sl(2,\CC)$ in terms of the two
copies of the R-deformed Heisenberg algebra. The spinor set
of relativistic equations based on this representation is
considered in Section \ref{eqns}, where we show that for
the finite-dimensional representations of RDHA such
equations universally describe massless states
with any integer and half-integer helicity. The
infinite-dimensional representations correspond formally to
the states with fractional helicities. We demonstrate that
such solutions, however, break down the Poincar\'e
invariance. In Section \ref{anyons} we show that the
compactification of the initial massless equations and
subsequent reduction to (2+1)-dimensional space result in
the consistent theory
of massive anyons. The absence of the massless
infinite-dimensional representations of the (3+1)$D$
Poincar\'e group is proved in Section \ref{nogo} in a
generic case, independently on the concrete form of
equations. Section \ref{rem} is devoted to a brief
discussion of the obtained results. Appendix reviews
briefly the Dirac equations for massive spinless positive
energy states.

\section{RDHA and $sl(2,\CC)$: generalization of the
Schwinger construction}
\label{rdha}

The first (3+1)$D$ relativistic equation, due to which
the infinite-dimensional unitary representations of
$SL(2,\CC)$ were discovered, is the Majorana equation
\cite{m}. Its solutions, however, describe {\it reducible}
representations of $ISO(3,1)$ characterized by
the positive energy in the massive sector $p^2<0$.
At the beginning of 70s Dirac \cite{dirac,dirac1}
(see also \cite{stau}) proposed a covariant spinor
set of equations (see Appendix) from which the Majorana
and Klein-Gordon equations appear in the form of
integrability conditions. As a result, the Dirac spinor set
of equations possesses a massive spin-$0$ positive energy
solutions, whereas its vector modification considered
by Staunton \cite{stau} describes massive spin-$1/2$
states\footnote{More details on infinite component
relativistic equations and corresponding $SL(2,\CC)$
representations can be found in Ref. \cite{st,bog}.}.
We are interested in analysing the possibility of
constructing relativistic wave equations for a
massless field carrying fractional helicity. For the
purpose, the fields related to the infinite-dimensional
representations of $sl(2,\CC)$ will be considered.

\subsection{Fock representations of the R-deformed algebra}

Having in mind the analogy with the (2+1)$D$
case of anyons, it is convenient to use the
infinite-dimensional representations of $sl(2,\CC)$ realized
by means of the two copies of RDHA
\cite{yang,rdha,rdha0,spin1}
with mutually commuting generators\footnote{The necessary
infinite-dimensional half-bounded representations of
$sl(2,\CC)$ \cite{barg,ggv} can be realized alternatively in
terms of homogeneous monomials \cite{flie} but the RDHA
construction is more convenient for our purposes.}:
\begin{equation}\label{eq:rdha}
 \begin{array}{lll}
 \left[a^-,a^+ \right] =\Pi + \nu R, \qquad\qquad&
 \{R,a^\pm\}=0, \qquad\qquad& R^2=\Pi,\\[-1mm]\\[-1mm]
 \left[\bar a^-, \bar a^+ \right]=\bar\Pi +\nu\bar R,&
 \{\bar R,\bar a^\pm\}=0,& \bar R^2=\bar\Pi.
\end{array}
\end{equation}
The operators $a^\pm$ will represent internal (spin)
degrees of freedom. They generalize the two
oscillator degrees of freedom (with $\nu=0$)
used by Dirac \cite{dirac,dirac1}.

In the case of a direct sum of representations of the
algebras with which we begin our analysis,
$\Pi$ and $\bar\Pi$ are the projectors on the
corresponding subspaces that in a matrix realization
is reflected by the relations $\Pi=\frac 12(1+\sigma_3)$,
$\bar\Pi=\frac 12(1-\sigma_3)$.
The operators $R$, $\bar R$ have the sense of
reflection operators for the internal variables
and $\nu\in\RR$ is the deformation parameter.
As it was mentioned above, the RDHA and its representations
were studied extensively in the literature. Here we will
mainly refer to \cite{rdha0}, where a universality of
the RDHA was observed: when $\nu =-(2k+1)$,
$k\in\NN$, its representations are finite-dimensional
(parafermion-like), and are infinite-dimensional if not
(being unitary paraboson-like for $\nu>-1$ \cite{rdha}).
The choice $\nu=0$ with a direct product of representations
of the two algebras was used in the Dirac
\cite{dirac,dirac1} and Staunton \cite{stau} sets of
equations.

For the sake of clarity and self-contained presentation,
we recall briefly the construction of representations of the
algebra. The infinite-dimensional representations are
built from the primitive vectors
$|0\rangle$ and ${|\bar 0\rangle}$ (vacua), annihilated by
$a^-$ and $\bar a^-$, respectively,
\begin{equation}\label{eq:rep_rdha}
 {\cal R}_\nu=\{|n\rangle=\frac{\left(a^+\right)^n}
 {\sqrt{|[n]_\nu!|}}|0\rangle,\ n\in\NN\}, \qquad
 \bar{\cal R}_\nu =\{|\bar n\rangle=
 \frac{\left(\bar a^+\right)^n}{\sqrt{|[n]_\nu!|}}
 |\bar 0\rangle, \ n\in\NN\},
\end{equation}
with $[n]_\nu!=[n]_\nu[n-1]_\nu\cdots[1]_\nu$,
$[0]_\nu!=1$, $[n]_\nu =n+\frac 12(1-(-1)^n)\nu$.
The finite-dimensional representations of RDHA with
$\nu=-(2k+1)$, $k\in\mathbb N$, are built
in the same vein (\ref{eq:rep_rdha}) but in this case
there is another primitive vector $|2k\rangle$ annihilated
by $a^+$, $a^+|2k\rangle=0$ \cite{rdha0}.

Let us consider the quadratic operators
\begin{equation*}
 \begin{array}{llll}
 J_\pm = \frac{1}{2} (a^\pm)^2,\qquad&
 J_0=\frac{1}{4}\{a^+,a^-\},\qquad&
 \bar J_\pm = \frac{1}{2} (\bar a^\pm)^2,\qquad&
 \bar J_0=\frac{1}{4}\{\bar a^+,\bar a^-\},
 \end{array}
\end{equation*}
forming the two copies of $sl(2,\RR)$ algebra,
$$
 [{\cal J}_0,{\cal J}_\pm]=\pm {\cal J}_\pm,\qquad\qquad
 [{\cal J}_-,{\cal J}_+]=2{\cal J}_0,
$$
where ${\cal J}=J$ or $\bar{J}$, and $[J,\bar{J}]=0$.
As a result, irreducible representations ${\cal R}_\nu$
and $\bar {\cal R}_\nu$ (\ref{eq:rep_rdha}) are decomposed
into the direct sums of the irreducible representations of
$sl(2,\RR)$ bounded from below as \cite{rdha0}
\begin{eqnarray}
\label{eq:rdha_lorentz}
{\cal R}_\nu = {\cal D}_{{\frac{1+\nu}{4}}}^+
\oplus{\cal D}_{{\frac{3+\nu}{4}}}^+, \qquad\qquad
\bar {\cal R}_\nu = \bar{\cal D}_{{\frac{1+\nu}{4}}}^+
\oplus\bar{\cal D}_{{\frac{3+\nu}{4}}}^+,
\end{eqnarray}
where $j_0=\kappa+l$, $l=0,1,\ldots$, are
the eigenvalues of ${\cal J}_0$
with $\kappa=\frac{1+\nu}{4}$ and
$\kappa=\frac{3+\nu}{4}$, respectively.
The ``left''- and ``right''-handed
parts of the generators of the
(3+1)$D$ Lorentz algebra
$sl(2,\CC)$, $K_i$ and $\bar{K}_i$, $i=1,2,3$,
obeying the relations
$$
 [{\cal K}_i,{\cal K}_j]=i\epsilon_{ijk}{\cal K}_k,
 \qquad\qquad[K,\bar{K}]=0,
$$
with ${\cal K}=K$ ($\bar{K}$),
are defined in terms of the $so(3,1)$ generators
$J_{\mu\nu}$ as follows:
$$
 K_i=\frac{1}{2}\epsilon_{ijk}J_{jk}+iJ_{0i},\qquad\qquad
 \bar{K}_i=\frac{1}{2}\epsilon_{ijk}J_{jk}-iJ_{0i}.
$$
Then the $sl(2,\CC)$ generators ${\cal K}_i$
can be identified with the $sl(2,\RR)$
generators ${\cal J}_0$,
${\cal J}_\pm={\cal J}_1\pm i{\cal J}_2$
by means of the relations
\begin{equation} \label{j2k}
\begin{array}{lll}
 {\cal J}_0=-{\cal K}_2,\qquad\qquad&
 {\cal J}_1=-i{\cal K}_1,\qquad\qquad&
 {\cal J}_2=-i{\cal K}_3.
\end{array}
\end{equation}
Such identification corresponds to the concrete choice
of the $\gamma$-matrices in covariant spinor formalism
(see below). Any other possible identifications between
$\cal J$ and $\cal K$, e.g., those obtained from (\ref{j2k})
by a cyclic permutations of ${\cal K}_i$, lead to other
realizations of the $\gamma$-matrices related by unitary
transformations.

Note here that the Fock spaces of the usual oscillators
corresponding to $\nu=0$ are decomposed into the spin-$1/4$
and spin-$3/4$ representations of $sl(2,\RR)$
\cite{perelomov}.
Another realization of $sl(2,\RR)$ generators,
$J_\pm = \frac{1}{2} (a^\mp)^2$,
$J_0=-\frac{1}{4}\{a^+,a^-\}$,
results in the direct sum of bounded from above
infinite-dimensional representations,
${\cal R}_\nu={\cal D}^-_{\frac{1+\nu}{4}}
\oplus {\cal D}^-_{\frac{3+\nu}{4}}$.

Thus, we have generalized the Schwinger construction of
$su(2)$ to the case of the $sl(2,\CC)$ algebra.

\subsection{Covariant spinor formalism}

Introducing $sl(2,\CC)$ notations of dotted and
undotted indices for two-dimensional spinors,
all can be rewritten in covariant notations. The spinor
conventions to raise/lower indices are as follow \cite{wb}:
$\psi_\alpha =\varepsilon_{\alpha\beta}\psi^\beta$,
$\psi^\alpha =\varepsilon^{\alpha\beta}\psi_\beta$,
$\bar\psi_{\dot\alpha}=\varepsilon_{\dot\alpha
\dot\beta}\bar\psi^{\dot\beta}$, $\bar\psi^{\dot\alpha}
=\varepsilon^{\dot\alpha\dot\beta}\bar\psi_{\dot\beta}$
with $(\psi_\alpha)^* =\bar\psi_{\dot\alpha}$,
$\varepsilon_{12} = \varepsilon_{\dot 1\dot 2}=-1$,
$\varepsilon^{12} = \varepsilon^{\dot 1\dot 2}=1$.
Defining the two spinor operators
$L_\alpha$ and $\bar L_{\dot{\alpha}}$,
\begin{equation}\label{eq:osp1|4spin}
\begin{array}{ll}\ds
 L_1= \frac{1}{\sqrt{2}} \left(a^+ + a^- \right),
 \qquad\qquad&\ds
 L_2= \frac{i}{\sqrt{2}} \left(a^+ - a^- \right) ,
 \\[-1.5mm]\\[-1.5mm]\ds
 \bar L_{\dot 1}=\frac 1{\sqrt 2}\left(
 \overline a^+ + \overline a^-\right),
 \qquad\qquad&\ds
 \bar L_{\dot 2}=\frac i{\sqrt 2}\left(
 \overline a^+ - \overline a^- \right),
\end{array}
\end{equation}
a direct calculation shows that they
generate the
$osp(4|1)$ superalgebra. Its bosonic part is
$sp(4,\RR)\sim so(3,2) = AdS_4$ with generators
\begin{equation}\label{eq:osp1|4}
\begin{array}{lll}\ds
 L_{\alpha\beta}=\frac 14\left\{L_\alpha,L_\beta\right\},
 \qquad\qquad&\ds
 L_{\dot\alpha\dot\beta}=\frac 14\left\{L_{\dot\alpha},
 L_{\dot\beta}\right\},
 \qquad\qquad&\ds
 M_{\alpha \dot \alpha } = \frac{1}{ 4} \left\{L_{ \alpha},
 L_{\dot \alpha} \right\},
\end{array}
\end{equation}
but we shall be interested only in its $so(3,1)$ part.
Let us introduce the $4D$
Dirac matrices in the Weyl representation,
\begin{eqnarray}
\label{eq:gamma}
\gamma_\mu =
 \begin{pmatrix}
 0&\sigma_\mu\\
 \bar\sigma_\mu&0
 \end{pmatrix},
\end{eqnarray}
with
\begin{eqnarray}\label{eq:spin_rep}
 \sigma_{\mu\,\alpha\dot\alpha}=\Bigl(1,\sigma_i \Bigr),
 \qquad\qquad\bar\sigma_\mu{}^{\dot\alpha\alpha}=
 \Bigl(1,-\sigma_i\Bigr).
\end{eqnarray}
Then the $so(3,1)$ spinor generators
$$
 \gamma_{\mu \nu}= \frac{i}{4} [\gamma_\mu,\gamma_
 \nu]=\mathop{\rm diag}\left(i\sigma_{\mu\nu}\,_\alpha
 {}^\beta,\ i\bar\sigma_{\mu\nu}\,^{\dot\alpha}
 {}_{\dot\beta}\right)
$$
allow us to present the $so(3,1)$ generators in the form
\begin{eqnarray}
\label{eq:so1,3}
 J_{\mu \nu}=\frac{1}{2}\left( L^{\alpha }
 \sigma _{\mu \nu\,\alpha }{}^{\beta }
 L_{\beta}-\bar{L}_{ \dot{ \alpha}}\bar{\sigma}_{\mu
 \nu}{}^{\dot{\alpha}}{}_{\dot{\beta}}
 \bar{L}^{\dot{\beta}}\right).
\end{eqnarray}
The Pauli-Lubanski pseudo-vector is given by
\begin{eqnarray}\label{eq:pauli-lub}
 W^\mu=\frac 12\varepsilon^{\mu\nu\rho\sigma}
 P_\nu J_{\rho\sigma}
\end{eqnarray}
with $P_\mu$ being the generator of space-time
translations.

\subsection{Invariant scalar product}

For any physically admissible representation of
$ISO(3,1)$, the generators of the Lorentz group
should be Hermitian. This means that for any such a
representation we have to construct an invariant
scalar product with the necessary properties. In what
follows we will discuss mainly the representation
${\cal D}^+_\lambda\oplus\bar{\cal D}^+_\lambda$.
Therefore, we consider in detail the
construction of the invariant scalar product for this
representation only. Though for free massless states the
left- and right-handed sectors are uncoupled, the
both are needed for the construction of the
invariant scalar product \cite{siegel}. We
consider the vectors living on the reducible
representation space,
$\Psi\in{\cal R}_\nu\oplus\bar{\cal R}_\nu$, i.e.
$\Psi=|\psi\>+|\bar\chi\>$ with $|\psi\>\in{\cal R}_\nu$ and
$|\bar\chi\>\in\bar{\cal R}_\nu$.

The representations of (\ref{eq:rdha}) possess the
natural involution
\begin{equation*}
\begin{array}{llll}
 \left(a^\pm\right)^+=a^\mp,\qquad&
 |0\rangle^+=\langle 0|,\qquad&
 \left(\bar a^\pm\right)^+=\bar a^\mp,\qquad&
 |\bar 0\rangle^+=\langle\bar 0|.
\end{array}
\end{equation*}
This involution is not a covariant operation since it
does not mix the left- and right-handed sectors. As a
consequence, the state $\langle\psi^*|=|\psi\rangle^+$
is not contravariant while the original state
$|\psi\rangle$ is a covariant vector of the
representation space. The Lorentz
generators are not Hermitian with respect to such an
involution. In order to obtain the covariant
(Hermitian) conjugation, we introduce the
intertwining operator $\Upsilon$ which permutes the
left- and right-handed sectors,
\begin{equation*}
\begin{array}{lllll}
 \Upsilon a^\pm=\bar a^\pm\Upsilon,\qquad&
 \Upsilon R=\bar R\Upsilon,\qquad&
 \Upsilon|0\rangle=|\bar 0\rangle,\qquad&
 \Upsilon |\bar 0\rangle=|0\rangle,\qquad&
 \Upsilon^2=1.
\end{array}
\end{equation*}
For the finite-dimensional representation
$\left(\frac 12,0\right)\oplus\left(0,\frac 12\right)$, the
matrix elements of this operator correspond to the
usual $\gamma^0$-matrix \cite{siegel}. In general this
operator can be represented as
$$
 \Upsilon =\sum_n(|\bar{n}\rangle
 \langle n|+|n\rangle \langle \bar{n}|).
$$

The state $\langle\bar\psi|\equiv\langle\psi^*|\Upsilon$ is
a contravariant vector of the representation. Therefore,
the $sl(2,\CC)$ invariant scalar product is
$\Psi^\dag\Psi$, where the Dirac-like conjugation is
defined by $\Psi^\dag =\Psi^+\Upsilon$. One
can verify that the Lorentz operators are Hermitian
with respect to this scalar product,
\[
 J_{\mu\nu}^\dag =\Upsilon J_{\mu\nu}^+\Upsilon
 =J_{\mu\nu}.
\]
For this conjugation we also have
$(L_\alpha)^\dag =\bar L_{\dot\alpha}$
and $R^\dag=\bar{R}$.

The quantum mechanical (positively
defined)
probability density
$\langle\psi^*|\psi\rangle =\langle\bar\psi|\Upsilon|
\psi\rangle$ (the left-handed part here)
is not a covariant object. For example, for the
above mentioned finite-dimensional representation it
corresponds to the zero component of the vector current:
$\bar\psi\gamma^0\psi$, where $\psi$ is the usual Dirac
spinor.

\section{Relativistic spinor equations for massless
states}
\label{eqns}

In this section we propose and investigate relativistic
spinor equations of covariant form based on the
representation of $sl(2,\CC)$ realised in terms of RDHA. It
turns out that from the algebraic point of view they
correspond to massless states of arbitrary real helicity.
For the finite-dimensional representations of RDHA the
equations universally describe states with any integer and
half-integer helicity. However, the application of the
infinite-dimensional representations of RDHA inevitably
leads to a Lorentz symmetry breaking.

\subsection{Algebraic aspect of equations}

Using definitions and conventions of the previous section,
let us introduce the fields
$|\psi\rangle \in{\cal R}_\nu$,
$|\bar\psi\rangle\in{\cal\bar R}_\nu$.
By analogy with the (2+1)$D$ case \cite{spin,spin1} we
postulate the relativistic wave equations
\begin{equation}\label{eq:spin}
 P^\mu\bar\sigma_\mu{}^{\dot\alpha\alpha}L_\alpha
 |\psi\rangle=0, \qquad\qquad
 P^\mu \sigma_{\mu\,\alpha\dot\alpha}\bar L^{\dot\alpha}
 |\bar\psi\rangle=0.
\end{equation}

Let us analyse the physical content of the equations
(\ref{eq:spin}) from the algebraic point of view.
Using the identity
\begin{eqnarray}
 \label{eq:spin-id}
 \Bigl(P^\mu\bar\sigma_\mu{}^{\dot\alpha\alpha}
 L_{\alpha}\Bigr)
 \Bigl(P^\nu\bar\sigma_\nu{}^{\dot\beta\beta}
 L_\beta\Bigr)
 \varepsilon_{\dot\alpha\dot\beta}
 =iP^\mu P_\mu (1 + \nu R),
\end{eqnarray}
and similar one for the right-handed part, we arrive at
the equations
\begin{equation}\label{eq:massless}
 P^\mu P_\mu |\psi\rangle=0,\qquad\qquad
 P^\mu P_\mu |\bar\psi\rangle=0,
\end{equation}
and, as a consequence, the proposed relativistic
equations describe relativistic massless fields.
Moreover, it is easy to see that
the solutions to the equations
(\ref{eq:spin}) obey also the relations
\begin{equation}\label{hel}
 \left(W^\mu+\tfrac{1+\nu}4P^\mu\right)
 |\psi\rangle=0, \qquad\qquad
 \left(W^\mu-\tfrac{1+\nu}4P^\mu\right)
 |\bar\psi\rangle=0.
\end{equation}
This means that they have the fixed helicity
$\lambda=-\frac{1+\nu}4$ for the left-handed sector and
$\lambda=\frac{1+\nu}4$ for the
right-handed one. So, from the viewpoint of the
Poincar\'e algebra the equations (\ref{eq:spin}) describe
massless states with arbitrary helicity.

In what follows we mainly concentrate on the left-handed
part since all the corresponding results for the
right-handed sector can be reproduced straightforwardly. In
components the first equations in (\ref{eq:spin}) read
\begin{equation}
\begin{array}{c}
 \Bigl(a^{+}\left( P^{0}+iP_{1}-P_{2}+P_{3}\right)
 -a^{-}\left ( P^{0}-iP_{1}+P_{2}+P_{3}\right)\Bigr)
 |\psi\rangle =0,
 \\[-2mm]\label{eq:spin1}\\[-2mm]
 \Bigl(a^{+}\left( P^{0}-iP_{1}-P_{2}-P_{3}\right)
 +a^{-}\left ( P^{0}+iP_{1}+P_{2}-P_{3}\right)\Bigr)
 |\psi\rangle =0.
\end{array}
\end{equation}
Taking the sum and the difference of these equations,
we get
\begin{equation}
\begin{array}{c}
 \Bigl(a^{+}(P^{0}-P_{2})+a^{-}\left( iP_{1}-P_{3}
 \right)\Bigr) |\psi\rangle = 0,
 \\[-2mm]\label{eq:spin2}\\[-2mm]
 \Bigl(a^{-}(P^{0}+P_{2})-a^{+}\left( iP_{1}+P_{3}
 \right) \Bigr) |\psi\rangle = 0.
\end{array}
\end{equation}

\subsection{Finite-dimensional representations: integer and
half-integer helicities}

Let us first consider a solution to the equations
(\ref{eq:spin2}) in the case of the finite-dimensional
representations of RDHA \cite{rdha0} with
$\nu=-(2k+1)$, $k\in\mathbb N$. As was noted above, the
finite-dimensional representations are built similarly to
(\ref{eq:rep_rdha}) and are characterised by the additional
primitive vector $|2k\rangle$ annihilated by
$a^+$, $a^+|2k\rangle=0$. Such representations of RDHA are
also reducible with respect to the $sl(2,\RR)$ algebra,
${\cal R}_\nu={\cal D}_{\frac k2}\oplus
{\cal D}_{ \frac{k-1}2}$.

It is worth noting that formally the solutions to the
equations (\ref{eq:spin2}) have the structure similar to
that of the Weyl equation $p^\mu\sigma_\mu\psi$=0 written
in components as
\begin{equation*}
\begin{array}{c}
 (p^0+p_3)\psi_1+(p_1-ip_2)\psi_2=0, \\[1mm]
 (p_1+ip_2)\psi_1+(p^0-p_3)\psi_2=0.
\end{array}
\end{equation*}
Indeed, the cone $\Gamma =\{p_\mu:p_\mu p^\mu=0,\
p^0\ne 0\}$ can be covered by the two charts
$U_\pm=\{p_\mu: p^0\pm p_3\ne 0\}$.
In each chart the solution can be represented in the
regular form
\begin{equation}\label{weyl}
 \psi(p)\Bigr|_{U_+}=
 \begin{pmatrix}\omega(p)\\1\end{pmatrix}
 \varphi_+(p)\delta(p^2), \qquad\qquad
 \psi(p)\Bigr|_{U_-}=
 \begin{pmatrix}1\\\tilde\omega(p)\end{pmatrix}
 \varphi_-(p)\delta(p^2),
\end{equation}
where the functions
$$
\omega(p)=\frac{ip_2-p_1}{p^0+p_3},\qquad\qquad
\tilde\omega(p)=\frac{p_1+ip_2}{p_3-p^0}
$$
obey on the cone the identity
$\omega(p)\tilde\omega(p)=1$. On the intersection
$U_+\cap U_-$, the functions $\varphi_\pm$ are related as
$\varphi_-(p)=\omega(p)\varphi_+(p)$.

The solution to the equations (\ref{eq:spin2}) can be
considered in the same way but with the covering
charts $U_\pm=\{p_\mu: p^0\pm p_2\ne 0\}$. The peculiarity
of the $p_2$-direction is associated with the chosen
representation of the $\sigma$-matrices and can be changed
to any other direction by a unitary transformation. The
solution to the first equation from (\ref{eq:spin}) is
$$
 |\psi\rangle\Bigr|_{U_+}=\delta(p^2)\varphi_+(p)
 \sum_{n=0}^{k}C_n\Omega^n(p)|2n\rangle,
 \qquad
 |\psi\rangle\Bigr|_{U_-}=\delta(p^2)\varphi_-(p)
 \sum_{n=0}^{k}C_n\tilde\Omega^{k-n}(p)|2n\rangle,
$$
with the functions $\varphi_\pm$ related on
$U_+\cap U_-$ as
$\varphi_-(p)=\Omega^k(p)\varphi _+(p)$.
Here $\varphi_\pm$ are the functions regular on
$U_\pm$, and
\begin{equation}\label{Om-Cn}
 \Omega(p)=\frac{p_3+ip_1}{p^0+p_2},\qquad\qquad
 \tilde\Omega(p)=\frac{p_3-ip_1}{p^0-p_2},\qquad\qquad
 C_n=\frac{\sqrt{|[2n]_\nu!|}}{2^nn!},
\end{equation}
with the identity $\Omega(p)\tilde\Omega(p)=1$
to be valid on the cone. {}From the explicit form of the
solution one can see that the equations of motion
(\ref{eq:spin}) contain effectively the projector on the
even invariant subspace ${\cal D}_{\frac k2}$ (or
$\bar{\cal D}_{\frac k2}$ for the right-handed sector). Here
we imply that the parity is defined with respect to the
action of the operator $R$ ($\bar R$). The obtained solution
describes a free left-handed massless particle with helicity
$\frac k2$. For example, in the case of helicity
$\frac 12$ the corresponding solution is given by
$$
 |\lambda=\tfrac 12\rangle\Bigr|_{U_+}=
 \delta(p^2)\varphi_+(p)
 (|0\rangle+\Omega(p)|2\rangle),
 \qquad
 |\lambda=\tfrac 12\rangle\Bigr|_{U_-}=
 \delta(p^2)\varphi_-(p)
 (|0\rangle+\tilde\Omega(p)|2\rangle).
$$
This solution is in the exact correspondence with the
solution (\ref{weyl}) to the (unitarily transformed) Weyl
equation. The solution to the second equation from
(\ref{eq:spin}) can be considered analogously.

Thus, {\em in the case of the finite-dimensional
representation, the equations (\ref{eq:spin}) provide a
consistent \underline{\em universal} description of the free
states with arbitrary integer and half-integer helicities.
The equations have the same form for all such helicities and
the information about the values of $\lambda$
is encoded in the parameter $\nu$.}

\subsection{Infinite-dimensional representations:
Lorentz symmetry\protect\\ breaking}

The equations (\ref{eq:spin}) formally have covariant
form and, as we have seen, give the consistent description
of the massless finite-dimensional representations of the
Poincar\'e group. But the situation for the
infinite-dimensional representations turns out to be
essentially different. A simple analysis of equations
(\ref{eq:spin}) reveals some contradictory properties of the
solutions: they exist in some frames and do not exist in
others. Indeed, the solution in the frame where
$p^\mu=(E,0,E,0)$ is given by $\psi\propto|0\rangle$,
with $|0\rangle$ the vacuum state of the RDHA.
However, no normalized solutions can be found in the frames
where $p^\mu=(E,0,-E,0)$, $p^\mu=(E,\pm E,0,0)$ or
$p^\mu=(E,0,0,\pm E)$. This means that {\em the solutions
are not invariant under some transformations of the
Lorentz group, and so, the Lorentz invariance is
broken.} Note, that if another choice of the Dirac matrices
would have been done, the Lorentz invariance would be
broken in other directions. We would like to note that this
situation with the Lorentz breaking has a formal analogy
with the usual spontaneous breaking of a global symmetry in
field theory models, where the equations of
motions are covariant with respect to the corresponding
symmetry group and the breaking occurs on the level of
vacuum solutions.

To make the statement on the breaking of the Lorentz
invariance more precise, we observe that
for the covering $\{U_+,U_-\}$ of the cone the formal
solution to the first equation from (\ref{eq:spin})
exists on $U_+$ only,
\begin{equation}\label{sol}
 |\psi\rangle =\delta(p^2)\varphi(p)
 \sum_{n=0}^{\infty}C_{n}\Omega^n(p)|2n\rangle=
 \delta(p^2)\varphi(p)\exp\left(\tfrac 12\Omega(p)
 (a^+)^2\right)|0\rangle,
\end{equation}
where $\varphi(p)$ is a regular function, and
$C_{n}$ and $\Omega(p)$ were defined in (\ref{Om-Cn}).
Some care has to be taken with infinite-dimensional
representations since, generally, an infinite-dimensional
representation of the Poincar\'e algebra is not
obligatory a representation of the Poincar\'e group.
Therefore, the solution (\ref{sol}) is proper one if its
norm with respect to the internal space scalar product,
\[
 \langle\psi^*|\psi\rangle=\sum_{n=0}^{\infty}
 C_{n}^{2}\left|\Omega(p)\right|^{2n},
\]
is finite. The radius of convergence of the series is equal
to
$$
 \lim_{n\to\infty}\frac{C_{n+1}}{C_{n}}=1.
$$
Therefore, we arrive at the strict inequality
$|\Omega(p)|^2<1$, which can be rewritten as
\begin{equation}\label{eq:lb}
 p_2\left(p^{0}+p_{2}\right)>0.
\end{equation}
All the formulas corresponding to the right-handed
sector can be reproduced via the formal substitution
$a^\pm\to\bar a^\pm$, $p_2\to-p_2$. Therefore, the
corresponding convergence condition for that sector is
\begin{equation}\label{eq:lb2}
 p_{2}\left(p^{0}-p_{2}\right)<0.
\end{equation}
Consequently, although the equations (\ref{eq:spin}) look
like $SL(2,\CC)$-invariant equations, the maximal invariance
group of the solution is the one preserving the
inequality (\ref{eq:lb}) (or (\ref{eq:lb2}) for
the right-handed sector). This means that
the equations are invariant under $SL(2,\RR)$ subgroup of
$SL(2,\CC)$ (the group generated by the rotations in
the plane (1-3) and by the boosts in the directions 1 and
3, which do not violate the relations (\ref{eq:lb}),
(\ref{eq:lb2})). This is a direct consequence of the fact
that the used infinite-dimensional representation is ill
defined at the level of the Poincar\'e group, even if it is
perfectly defined at the level of the Lie algebra.

The solution (\ref{sol}) illustrates an unusual type
of Lorentz symmetry breaking. The massless equations are
formally covariant but in the case of infinite-dimensional
representations the Lorentz invariance is strongly broken on
the level of the solutions. Formally, this breaking is
associated not with the infinite-dimensional representation
of the $SL(2,\CC)$ group itself but rather with the attempt
to enclose it into the corresponding representation of the
Poincar\'e group (cf. with massive case, see Appendix).
Indeed, as follows from the inequalities (\ref{eq:lb}),
(\ref{eq:lb2}), the Lorentz violation is provoked by the
transformations of the momentum, which is naturally
associated with the Poincar\'e group.

Through the reduction $sl(2,\CC)\longrightarrow sl(2,\RR)$,
the infinite-dimensional representations ${\cal R}_\nu$ and
$\bar{\cal R}_\nu$ are identified. In this case the
representation ${\cal D}_\lambda^\pm$ can be
exponentiated to a representation of the Lie group
$ISO(2,1)$. So, the Poincar\'e invariance in
(3+1)$D$, $ISO(3,1)$, is broken to
$ISO(2,1)$, the Poincar\'e invariance in (2+1)$D$.
As we shall see in the next section, this means that the
dimensional reduction to the direction $p_2$ gives rise to a
consistent (2+1)$D$ theory describing a massive anyon of
a mass $m$ and spin $\lambda=\pm\frac{1+\nu}4$.

{\em One has to stress once again that for the
case of the finite-dimensional representations, when
$2\lambda$ is an integer number, the problem encountered
with infinite-dimensional representations is not present
and, as we have seen, the consistent equations
for the helicity $\pm\lambda$ fields are obtained.}

\section{Compactification and reduction to (2+1)$D$ anyons}
\label{anyons}

We have seen that in the case of infinite-dimensional
representations the theory provided by the equations
(\ref{eq:spin}) does not have a nontrivial content on the
whole (3+1)$D$ Minkowski space $M^4$ since the formal
solution (\ref{sol}) breaks the Lorentz invariance. But let
us show that, in a sense, this problem can be ``cured'' by
compactifying the singled out (in this case $x_2$) direction
on a circle, $M^4\to M^3\times S^1$.

Fixing the space geometry in the form of the
three-dimensional Minkowski space times a circle of radius
$m^{-1}$ and denoting the compactified coordinate as
$\theta$, $0\le\theta\le 2\pi$, the state $|\psi\>$ in
coordinate representation can be expanded as
\begin{equation*}
 |\psi\>=\sum_{n=-\infty}^\infty e^{in\theta}|\psi_n\>,
 \qquad\text{with}\qquad
 |\psi_n\>=\sum_{k=0}^\infty\psi_{nk}(y^a)|k\>
\end{equation*}
depending only on the 3-dimensional coordinates $y^a$,
$a=0,1,2$, with $y^0=x^0$, $y^1=x^1$ and $y^2=x^3$.

So, starting with a (3+1)$D$ massless momentum $P^\mu$, it
is reduced to $(P^0,P^1, nm,P^3)$ for each state
$|\psi_n\>$. Denoting the three-dimensional momentum by
$\mathfrak p^a=(P^0,P^1,P^3)$, the mass-shell condition in
(2+1)$D$ reads
$$
 \left(\mathfrak p^a\mathfrak p_a+n^2m^2\right)|\psi_n\>=0,
$$
i.e. the initial
massless system is formally reduced to an infinite tower of
massive states plus a massless one. For $sl(2,\mathbb R)$
the left- and right-handed representations are equivalent,
and so, the dotted and undotted indices are identical. The
both equations in (\ref{eq:spin}), through the
compactification process, lead to
the equations
\begin{equation}\label{eq:spin2d}
 \left(\mathfrak p^a(\gamma_a)_\alpha{}^\beta
 +n m\delta_\alpha{}^\beta\right)L_\beta
 |\psi_n\rangle=0.
\end{equation}
These equations can be obtained from (\ref{eq:spin}) by the
formal substitution $P_2\to nm$ and multiplication by
$\sigma_2$ that, in fact, corresponds to lowering the
free index. The $\gamma$-matrices are given in the Majorana
representation:
\begin{equation*}
 (\gamma_0)_\alpha{}^\beta=-(\sigma_2)_\alpha{}^\beta,
 \qquad\qquad
 (\gamma_1)_\alpha{}^\beta=-i(\sigma_3)_\alpha{}^\beta,
 \qquad\qquad
 (\gamma_2)_\alpha{}^\beta=i(\sigma_1)_\alpha{}^\beta.
\end{equation*}
Up to the spatial rotation $y_1\to-y_2$, $y_2\to y_1$,
the equations (\ref{eq:spin2d}) actually coincide with the
equations for anyons \cite{spin,spin1} and, hence, for
$n\ne 0$ they {\it consistently} describe
relativistic (2+1)$D$ anyons of the spin
$\lambda=-\mathop{\rm sign}(n)\frac{1+\nu}4$.
By analogy with (\ref{sol}), the formal solutions to these
equations in the momentum representation are given by
\begin{equation}\label{eq:anyon}
 |\psi_n\rangle= \delta(\mathfrak p^2+n^2m^2)
 \varphi_n(\mathfrak p)\exp\left(\tfrac 12
 \Omega_n(\mathfrak p)(a^+)^2\right)|0\rangle,
\end{equation}
with
$\Omega_n(\mathfrak p)=\frac{\mathfrak p_2+i\mathfrak p_1}
{\mathfrak p^0+nm}$. The condition of normalizability
(\ref{eq:lb}) is transformed into the inequality
\begin{equation}\label{p0m}
 n(\mathfrak p^0+nm)>0.
\end{equation}
This means that the solutions (\ref{eq:anyon}) are
normalizable only in the case of energy with definite sign:
$\mathop{\rm sign}(\mathfrak p^0)=\mathop{\rm sign}(n)$.
In other words, all the states $|\psi_n\>$ with $n>0$ have
the positive energy, those with $n<0$ have negative energy
while the state $|\psi_0\>$ does not belong to the spectrum
of the theory. Such properties reveal the difference of
the considered compactification from the usual procedure.

Let us discuss this issue from viewpoint of the symmetries
usually associated with such a compactification on a circle.
Following Ref. \cite{duff}, one can infer that in the usual
case an affine Kac-Moody algebra $\frak g$ is always
associated with a compactified $M^3\times S^1$ theory. This
infinite-dimensional algebra consists of the loop extension
of $iso(2,1)$,
\begin{equation*}
 \widehat{iso}(2,1)=\{P_n^a=e^{in\theta}P^a, \
 J_n^a=e^{in\theta}J^a\},
\end{equation*}
where $P^a$ and $J^a$ are the translation and Lorentz
generators of $iso(2,1)$, respectively, and of the
additional set of operators
$Q_n=ie^{in\theta}\partial_\theta$ with the commutation
relations
\begin{equation*}
 [Q_n,Q_m]=(n-m)Q_{n+m},\qquad\quad
 [Q_n,P_m^a]=-mP_{n+m}^a,\qquad\quad
 [Q_n,J_m^a]=-mJ_{n+m}^a.
\end{equation*}
The affine Kac-Moody algebra $\frak g$ has the following
natural triangular decomposition
$$
 \frak g=\frak g_+\oplus\frak g_0\oplus\frak g_-,
$$
where the subalgebras $\frak g_+$ and $\frak g_-$ are the
sets of all the operators with $n>0$ and $n<0$,
respectively, while $\frak g_0=iso(2,1)\oplus u(1)$ with
$Q_0$ being the generator of the $u(1)$ algebra.

In our case the symmetry corresponding to the affine algebra
$\frak g$ is always partially broken.
Indeed, the action of $\frak g_-$ is ill defined on the part
of the spectrum with positive energy ($n>0$) while
$\frak g_+$ is ill defined on the part of the spectrum with
negative energy ($n<0$). Hence, the symmetry is always
broken to $\frak g_+\oplus\frak g_0$ or to
$\frak g_0\oplus\frak g_-$. The only symmetry of the whole
spectrum is $\frak g_0$. Nevertheless, one can conclude that
in spite of the partial breaking of the infinite-parametric
symmetry, the compactified theory has the non-trivial
content.

The considered compactification can be treated, in
principle, as that ``induced'' by the breaking of
the Lorentz invariance of the equations (\ref{eq:spin})
for the infinite-dimensional representations case.
The ``induction'' is understood here in the sense that the
compactified theory, unlike the initial one, is consistent
and the direction of the compactification is defined by the
Lorentz breaking.

The dimensional reduction emerges after choosing one level,
say $n=1$ or $n=-1$, and discarding all the others.
Evidently, the reduction gives rise to a consistent
(2+1)$D$ theory describing a massive anyon of a mass $m$ and
spin $\lambda=\pm\frac{1+\nu}4$. Correspondingly, the formal
$ISO(3,1)$ invariance is reduced to the true $ISO(2,1)$
invariance.

The equations (\ref{eq:spin2d}) providing solutions with
fixed sign energy have a close analogy with the massive
Dirac (3+1)$D$ positive energy spinor equations
\cite{dirac,dirac1}
(see Appendix)\footnote{The change of the
sign before mass in the Dirac equations leads to solutions
with negative energy.}.
The construction of the
positive energy massive Dirac equations is based, in fact,
on the representation of the type
${\cal R}_{\nu=0}\otimes\bar{\cal R}_{\nu=0}$,
and in the massless limit they are reduced to the
equations of the form (\ref{eq:spin}) to be imposed
on the same state. However, note that for $p^0>0$, the
inequalities (\ref{eq:lb}), (\ref{eq:lb2}) are incompatible.
This means that there is no reference frame in which the
normalizable solutions could exist in the left- and
right-handed sectors simultaneously. Using this observation,
one can assume that the Dirac equations \cite{dirac,dirac1}
have no proper solutions in the massless limit. In Appendix
we demonstrate that this is indeed the case.

\section{No-go theorem for massless infinite-dimensional
\protect\\ representations of $ISO(3,1)$}
\label{nogo}

We have seen that the proposed spinor sets of equations
cannot be used to describe a massless field with fractional
helicity in (3+1)$D$ because the maximal invariant group of
the solution is broken down to the (2+1)$D$ Poincar\'e
group. Moreover, the same problems arise under attempt to
describe the massless field of integer or half-integer
helicity by means of infinite-dimensional representations of
$sl(2,\CC)$ associated with RDHA. So, we may wonder if this
feature is general or it is specific to the equations
(\ref{eq:spin}) we have considered. In other words, formally
we may address the general problem {\it if fractional
helicity states in (3+1)$D$ might exist.} At this point, the
natural question we should ask concerns the other type of
equations that could be proposed. Following Ref. \cite{stau}
and the results obtained in (2+1)$D$ case \cite{vect}, a
vector set of equations can be considered,
\begin{equation}
 \label{eq:vect}
 (\alpha P^\mu+iJ^{\mu\nu}P_\nu)\Psi=0,
\end{equation}
where $\alpha$ is some constant that defines the
helicity while the representation of $\Psi$ is not
fixed here. According to Refs. \cite{siegel,siegel1}, the
equations of such a form describe all irreducible massless
finite-dimensional representations of the Poincar\'e group.
But we are going to consider these equations from the
viewpoint of infinite-dimensional representations. In this
sense the equations (\ref{eq:vect}) are analogous to the
massless limit of the equations proposed by Staunton in Ref.
\cite{stau}. Contraction with $P^\mu$ shows that we have a
massless field. Then, choosing the frame where
$P^\mu=E(1,0,\epsilon,0)$, we reduce the system of
equations to the system
\begin{eqnarray}
\label{eq:vect2}
 \Bigl(\alpha-\epsilon\left(J_0-\bar J_0\right)\Bigr)
 \Psi&=&0, \nonumber \\
 \Bigl((1+\epsilon)\left(J_-+\bar J_+\right)+
 (1-\epsilon)\left(J_++\bar J_-\right)\Bigr)\Psi&=&0,\\
 \Bigl((1+\epsilon)\left(J_--\bar J_+\right)-(1-\epsilon)
 \left(J_+-\bar J_-\right)\Bigr)\Psi
&=&0.
\nonumber
\end{eqnarray}
For $\epsilon=+1$, the solution has to obey the relations
\begin{equation}\label{bound1}
 J_-\Psi=\bar J_+\Psi=0,
\end{equation}
whereas for $\epsilon=-1$ they are changed for
\begin{equation}\label{bound2}
 J_+\Psi=\bar J_-\Psi=0,
\end{equation}
that is possible only for the finite-dimensional
representations of $sl(2,\CC)$. In other words, we arrive at
the same problems as before and Eqs. (\ref{eq:vect}) are not
consistent for the states carrying fractional helicity
or, generally, with all the infinite-dimensional
representations. On the other hand, one can show that the
dimensional reduction of (\ref{eq:vect}) leads to the
(2+1)$D$ vector set of equations that, as well as the spinor
equations (\ref{eq:spin2d}), {\em consistently} describes
anyons \cite{vect}.

On a general ground, any set of equations which
can be proposed to describe a massless field of helicity
$\lambda$ (fractional or not) has to give rise
to the ``helicity equation''
\begin{eqnarray} \label{eq:hel}
 \Bigl(W^{\mu}-\lambda P^{\mu}\Bigr)\Psi=0,
\end{eqnarray}
with $W^\mu$ the Pauli-Lubanski vector \cite{ryder}. The
representation of $\Psi$ is also not specified
providing the universality of the analysis.

The equations (\ref{eq:hel}), like eqs. (\ref{eq:spin}) and
(\ref{eq:vect}), are not compatible for
infinite-dimensional representations. Indeed, {\it e. g.},
in a frame where $P^{\mu}=E(1,0,\epsilon,0)$, the equations
are simplified for
\begin{eqnarray} \label{eq:hel4}
 \Bigl(\lambda -\epsilon \left(
 J_{0}+\bar{J}_{0}\right)\Bigr) \Psi &=& 0,
\nonumber \\
 \Bigl((1+\epsilon)\left(J_--\bar J_+\right)+(1-\epsilon)
 \left(J_+-\bar J_-\right)\Bigr)\Psi&=&0,
\\\nonumber
 \Bigl((1+\epsilon)\left(J_-+\bar J_+\right)-(1-\epsilon)
 \left(J_++\bar J_-\right)\Bigr)\Psi&=&0.
\end{eqnarray}
One can see that these equations can be obtained from
(\ref{eq:vect2}) by the substitution $\bar J_0\to-\bar J_0$,
$\bar J_\pm\to-\bar J_\pm$ and $\alpha\to\lambda$. The
reason is that the equations (\ref{eq:hel}) can be
reproduced from (\ref{eq:vect}) by the formal substitution
$J^{\mu\nu}\to i\tilde J^{\mu\nu}$, $\alpha\to\lambda$,
where $\tilde J^{\mu\nu}=\frac 12\epsilon^{\mu\nu\rho\sigma}
J_{\rho\sigma}$. For $\epsilon=1$ or $\epsilon=-1$ we
arrive, correspondingly, at the equations (\ref{bound1}) or
(\ref{bound2}). This means that the equations
(\ref{eq:hel}) are compatible for finite-dimensional
representations (with integer or half-integer helicities)
only. In other words, {\em the (3+1)D Poincar\'e
group has no massless irreducible representations of any
(integer, half-integer or fractional) helicity with the
trivial non-compact part of the little group constructed on
the basis of infinite-dimensional representations of
$sl(2,\CC)$.}

\section{Discussion and outlook}
\label{rem}

The specific properties of the (2+1)-dimensional
space-time admit the existence of anyons. The
little group of massive states in (2+1)$D$ coincides
with the compact part (which is the infinitely connected
group $SO(2)$) of the little group of massless
(3+1)$D$ states. Consequently, the ``charge'' of its
universal covering group $\overline{SO(2)}=\RR$ is not
quantized. The group theory justification of anyons is
related to the fact that the representations of the
fractional charge describing relativistic anyons can be
extended to representations of the whole (2+1)-dimensional
Lorentz group which is also infinitely connected. For
the purpose of describing the massless states with
fractional ($\lambda\in\mathbb R$) helicity we have
constructed representations of $sl(2,\CC)$ in terms of
two copies of the $R$-deformed Heisenberg algebra.
In terms of such
representations of the Poincar\'e algebra,
\begin{itemize}
 \item\it
 The universal relativistic equations (\ref{eq:spin})
 describing massless states of any integer and half-integer
 helicity have been proposed.
\end{itemize}
This possibility is related to the existence of the
finite-dimensional representations of RDHA for
$\nu={}-(2k+1)$, $k\in\mathbb N$. For $\nu>-1$, the RDHA
has infinite-dimensional unitary representations.
In this case the equations (\ref{eq:spin}) formally describe
the massless states with fractional helicity.
Analysing the solutions to the relativistic
wave equations, we have traced out explicitly that the
corresponding infinite-dimensional representations of the
Lie algebra $sl(2,\CC)$ cannot be exponentiated to
representations of the Lie group $SL(2,\CC)$ for such
massless states. In other words, the (3+1)$D$ Lorentz
invariance is broken on the level of solutions. This
resembles the Lorentz symmetry breaking in supersymmetric
Yang-Mills theory considered in Ref. \cite{nishino}, where
the corresponding action is Lorentz invariant while the
symmetry is broken at the field equation level. In spite of
the Lorentz symmetry breaking we have shown that
\begin{itemize}
 \item\it
 The dimensional reduction of the massless equations
 (\ref{eq:spin}) leads to the consistent (2+1)$D$ theory of
 massive anyons with spin $\lambda=\pm\frac{1+\nu}4$.
\end{itemize}
The dimensional reduction emerges from the compactification
$M^4\to M^3\times S^1$. We treat this compactification
as that induced by the Lorentz symmetry breaking in
the equations (\ref{eq:spin}) on the solution level. It
would be interesting to find an example of such a
compactification (induced by the Lorentz symmetry breaking
on solution level) in other theories.
We hope that the observed unusual symmetry breaking could be
helpful in the context of the considerable activity looking
for different mechanisms of the Lorentz symmetry violation
\cite{nishino,lb1,lb2,lb3,lb4,lb5,lb6,lb7}
and compactification \cite{sc1,sc2,sc3,sc4}.

Concluding our investigation on existence of massless
states with fractional helicity in the 4-dimensional
space, we have analysed the fundamental equation
(\ref{eq:hel}) defining massless irreducible representations
of $ISO(3,1)$, and have shown that
\begin{itemize}
 \item\it
 The Poincar\'e group $ISO(3,1)$ has no massless
 irreducible representations with the
 trivial non-compact part of the little group
 constructed on the basis of the infinite-dimensional
 representations of $sl(2,\CC)$.
\end{itemize}
This means, in particular, that {\em the massless
(3+1)-dimensional fractional helicity states cannot be
described in a consistent way and that the integer and
half-integer helicity massless particles can be described
only in terms of finite-dimensional representations of
$sl(2,\CC)$.} It is worth also emphasizing that the obtained
restriction on values of helicity is not of a topological
origin. The topological arguments \cite{y,wein} do not
forbid irreducible representations with integer and
half-integer helicity constructed on the basis of the
infinite-dimensional representations of $sl(2,\CC)$.
Therefore, {\em the topology provides only the necessary
condition for massless representations to be the true
representations of $ISO(3,1)$, which, as we see, is not the
sufficient condition.} This is quite an unexpected result
since the massive irreducible representations of $ISO(3,1)$
with integer and half-integer spin can be constructed on the
basis of the infinite-dimensional representations of
$sl(2,\CC)$ \cite{dirac,dirac1,stau}.

The situation with the equations (\ref{eq:spin}) for the
infinite-dimensional representation can be compared with the
result obtained earlier in Ref. \cite{pr} in a different
context. Here the proposed equations look invariant under
the (3+1)$D$ Poincar\'e transformations but relativistic
invariance is broken at the level of the solutions. In the
paper \cite{pr} two of us have obtained similar results,
which were related, however, to the fact that the considered
there Lorentz automorphism of the underlying algebra is
outer and not obligatory inner. Only when the outer
automorphism becomes inner, a covariant equation can be
obtained. All this means, in particular, that it is not
enough to have the equation (or the set of equations) which
looks invariant (or covariant) to obtain a covariant
theory.

\vskip 0.5cm
{\bf Acknowledgements}
\vskip 5mm

MP thanks L. Avarez-Gaume, A. I. Oksak, V. I. Tkach, G. G.
Volkov and A. A. Zheltukhin for early communications
stimulated the present research and acknowledges the useful
communications with L. N. Lipatov and L. Ryder. MRT
acknowledges gratefully the useful discussions with G.
Mennessier and M. J. Slupinski. One of us (MRT) would like
to thank USACH for its hospitality, where the part of this
work was realized. The work was supported in part by the
grants 1980619, 7980044, 1010073 and 3000006 from FONDECYT
(Chile) and by DICYT (USACH).

\appendix
\section{Dirac positive-energy relativistic equations}

The Dirac equations \cite{dirac,dirac1} corresponding to
massive spinless particle with positive energy can be
represented as
\begin{equation}\label{eq:dirac}
 \left(P^\mu\tilde\gamma_\mu-m\I\right)Q\Psi=0,
 \qquad\text{with}\quad Q=\begin{pmatrix}
 q_1\\q_2\\\eta_1\\\eta_2
 \end{pmatrix},
\end{equation}
where $q_1$ and $q_2$ are commuting dynamical quantities and
$\eta_1$ and $\eta_2$ denote the corresponding conjugate
momenta, $[q_i,\eta_j]=i\delta_{ij}$, $i,j=1,2$.
The matrices $\tilde\gamma_\mu$ are related to those in the
Weyl representation (\ref{eq:gamma}) by the unitary
transformation $\gamma_\mu=U\tilde\gamma_\mu U^\dag$ with
the matrix
\begin{equation*}
 U=\frac 1{\sqrt 2}\begin{pmatrix}
  0&0&1&i\\1&-i&0&0\\i&-1&0&0\\ 0&0&-i&-1
 \end{pmatrix}.
\end{equation*}

Dirac showed that the equations (\ref{eq:dirac})
consistently describe massive spin-0 positive-energy
states \cite{dirac,dirac1}.
In the ``coordinate'' representation, with the operators
$q_i$ to be diagonal, the solution to the equations
(\ref{eq:dirac}) has the form \cite{dirac,dirac1}
\begin{equation}\label{dsol}
 \Psi\propto\delta(p^2+m^2)\exp\bigg({}
 -\frac 1{2(p^0-p_3)}\bigl(m(q_1^2+q_2^2)
 -ip_1(q_1^2-q_2^2)+2ip_2q_1q_2\bigr)\bigg).
\end{equation}
One can verify that this solution is normalizable in
``internal'' variables for $m\ne 0$ only. Moreover, in the
massless limit the solution (\ref{dsol}) is singular on the
cone $p^2=0$ due to the presence of the factor
$(p^0-p_3)^{-1}$.
This means that the equations (\ref{eq:dirac}) have no
proper solution in the massless limit.

It is worth noting that the formal change $m\to-m$
leads to the normalized solutions with negative energy.


\end{document}